\documentclass[utf8]{FrontiersinHarvard} 

\usepackage{url,hyperref,lineno,microtype,subcaption}
\usepackage[onehalfspacing]{setspace}
\usepackage{times}
\usepackage{latexsym}
\usepackage{subcaption}
\usepackage{multirow,bigdelim}
\usepackage{amsmath}
\usepackage{rotating}
\usepackage{booktabs,xcolor}
\usepackage{xspace}
\usepackage{booktabs}
\usepackage{multirow}
\usepackage{todonotes}
\usepackage{url}
\usepackage{adjustbox}
\usepackage{listings}

\makeatletter
\@ifundefined{description}{

}{}
\makeatother

\newcommand{\red}[1]{\textcolor{red}{#1}}

\usepackage{etoolbox}
\makeatletter
\patchcmd{\@makefntext}{\insertfootnotetext}{\relax}{}{}
\makeatother
\def\keyFont{\fontsize{8}{11}\helveticabold }
\def\firstAuthorLast{Ive {et~al.}} 
\def\Authors{Julia Ive\,$^{1,2*,\dagger}$, Vishal Yadav\,$^{1,\dagger}$, Mariia Ignashina\,$^{1}$, Matthew Rand$^{1}$ and Paulina Bondaronek$^{2}$}
\def\Address{$^{1}$ Queen Mary University of London, School of Electronic Engineering and Computer Science, London, UK \\ }
\def\Address{$^{2}$ UCL Institute Of Health Informatics, 222 Euston Road, NW1 2DA, London, United Kingdom}
\def\Address{$^\dagger$ equal contribution \\ }
\def\corrAuthor{Julia Ive}

\def\corrEmail{julia.ive84@gmail.com}

\begin{document}

\onecolumn
\firstpage{1}

\title[Privacy-Preserving Behaviour of Chatbot Users]{Privacy-Preserving Behaviour of Chatbot Users: Steering Through Trust Dynamics} 

\author[\firstAuthorLast ]{\Authors} 
\address{} 
\correspondance{} 

\extraAuth{}

\maketitle

\begin{abstract}

\section{}
Introduction: The use of chatbots is becoming increasingly important across various aspects of daily life. However, the privacy concerns associated with these communications have not yet been thoroughly addressed. The aim of this study  was  to investigate user awareness of privacy risks in chatbot interactions, the privacy-preserving behaviours users practice, and how these behaviours relate to their awareness of privacy threats, even when no immediate threat is perceived. 

Methods: We developed a novel "privacy-safe" setup to analyse user behaviour under the guarantees of anonymization and non-sharing. We employed a mixed-methods approach, starting with the quantification of broader trends by coding responses, followed by conducting a qualitative content analysis to gain deeper insights. 

Results: Overall, there was a  substantial lack of understanding among users about how chatbot providers handle data (27\% of the participants) and the basics of privacy risks (76\% of the participants). Older users, in particular, expressed fears that chatbot providers might sell their data. Moreover, even users with privacy knowledge do not consistently exhibit privacy-preserving behaviours when assured of transparent data processing by chatbots. Notably, under-protective behaviours were observed among more expert users. 

Discussion: These findings highlight the need for a strategic approach to enhance user education on privacy concepts to ensure informed decision when interacting with chatbot technology. \red{This is critical for designing chatbots in the public health domain where protecting sensitive health information is required while maintaining user trust.} This includes the development of tools to help users monitor and control the information they share with chatbots. 
\tiny
 \keyFont{ \section{Keywords:} chatbot, privacy, privacy paradox, qualitative methods} 
\end{abstract}


\section{Introduction}

Chatbots are designed to simulate human-like conversations and provide automated responses to user queries. They have evolved significantly since their first successes in the 1960s~\citep{Weizenbaum1966} by incorporating the latest natural language processing advancements. This evolution includes such widely-used applications as Apple's Siri (\url{https://www.apple.com/siri}), Amazon's Alexa (\url{https://alexa.amazon.com}), culminating in such chatbots as ChatGPT (\url{https://chat.openai.com})\red{, Copilot (\url{https://copilot.microsoft.com}) and Gemini (\url{https://gemini.google.com}).}

Thanks to those successes, chatbots have become increasingly popular and widely used in various industries (financial~\citep{Lappeman2023}, legal~\citep{Willems2023}, public services~\citep{Larsen2024}, consumer~\citep{Dev2023} and medical (such as psychotherapy sessions)~\citep{Li2023}), and their rapid proliferation has raised a number of ethical concerns, such as trust, safety, bias, privacy, transparency, agency and accountability, as well as ecological issues related to the energy consumption. Those concerns raise the question on whether the benefits outweigh the risks. And it seems that this is the case for repetitive tasks such as scheduling and on-boarding in the legal domain~\citep{Rukadikar2024}. In the mental health domain, chatbots represent an attractive perspective of offering accessible help largely reducing the risk of stigmatisation~\citep{lgbt24}, but raising the issue of clinical outcome validity~\citep{Li2023}.  

The technology aspect of addressing those ethical concerns continues to evolve. For example by more sophisticated data curation procedures, as well as more advanced fine-tuning~\citep{Wei2023JailbrokenHD,Yu2024}. Personalisation plays the core role in building trust during the human-AI interaction: it makes interactions more sociable and relatable. This perceived social intelligence make users feel that the AI understands and adapts to them, so that they are more likely to trust its recommendations.~\citep{Araujo2024}. More and more attention is paid to the evaluation frameworks~\citep{wang2023decodingtrust,Gumusel24}. There are active attempts to create legal regulatory frameworks to address some of the ethics concerns. However, those frameworks have examined chatbot features without sufficient input from public service representatives and users, which limits their effectiveness in addressing ethical concerns comprehensively~\citep{worsdorfer24}.

Among the ethics issues mentioned above, privacy plays one of the pivotal roles: it is essential for building trust~\citep{veale}. Respecting privacy empowers agency and prevents manipulation, it also allows to control self-disclosure and promotes accountability.

Indeed, there are multiple aspects of chatbot interactions that threaten privacy. More ``natural'' conversations, perceived anthropomorphism of chatbots and nonverbal behaviour tend to increase self-disclosure~\citep{Ischen2020,Pizzi2023}. While interacting with users, chatbots gather substantial amounts of personal information, including voice recordings, text messages, and user preferences. This data collection raises concerns about how this information is stored, protected, and potentially shared with third parties. There is a risk of unintended data sharing when chatbots are prompted in a particular way when they can reveal the information memorised from the training data~\citep{unlearning}. Moreover, hackers and malicious actors may target chatbot systems to gain unauthorized access to user information, leading to identity theft, financial fraud, or other privacy breaches~\citep{Shah2022}. Furthermore, excessive sensitive information sharing can consequently increase the risk of manipulation from the side of chatbots. For example,~\cite{volker20} investigated the manipulation capacities of chatbots within personality-aware systems.

Users may not always be aware of the risks associated with sharing different types of sensitive information. This includes both direct and indirect identifiers, which pose identity disclosure risks, as well as personal facts that might make users uncomfortable. Additionally, there is often a lack of awareness about the extent to which their information is being collected and stored. For example, users may believe that ChatGPT searches and summarises the Internet to generate responses~\citep{Zhang2024}. Recent research highlights the importance of increasing the chatbot transparency. It also stresses the need for greater user consent and awareness regarding how their data are used~\citep{Gumusel24,Zhang2024}.

Such measures while reasonable, may not be sufficient, especially considering the fact that multiple mismatches between the user privacy concerns and privacy-seeking behaviours have been reported. For example, multiple studies report a phenomenon of the ``privacy paradox'' which states the discrepancy between information privacy attitude and actual behaviour, i.e., the more people are concerned about their privacy the more they tend to over-share their information~\citep{Kokolakis2017,Colnago2023} (see also the observations regarding the reverse version of the paradox, individuals who appear to overlook privacy concerns exhibit active privacy-seeking behaviour; along with attempts to disprove the existence this phenomenon~\citep{Solove2020}). The ``privacy paradox'' phenomenon seems to hold for the chatbot interactions (e.g., in the public services domain~\citep{Willems2023}), confirming that increasing the user awareness may not be enough. The privacy paradox holds strongly in chatbot interactions because of their high perceived usefulness, ease of interaction and anthropomorphic design leading users to prioritise short-term benefits over long-term privacy risks. 

Our study proposes a carefully designed ``privacy-safe'' setup which allowed us to analyse user behaviour in non-threatening low-risk conditions. The study uses an ad-hoc chatbot pre-programmed to ask variations of the same questions to the users using a script on three topics (graduation ceremony, morning routine and your pet) varying the type of sensitive information the answers could potentially contain. We investigate the awareness of privacy risks and protection measures relevant to human-chatbot communication, as well as privacy-preserving behaviours in the relation to this awareness. 

\red{The novelty of our ``privacy-safe'' setup lies in its departure from recent research trends. Most of the recent studies investigating disclosures rely on real-world conversations with commercial chatbots like ChatGPT~\citep{Gumusel24,Zhang2024}. In contrast, our setup avoids these risks entirely by using a controlled environment where conversations remain local, ensuring that no participant data is transmitted to third parties.  }

 
In particular, we seek to answer the following research questions:
\noindent 

(Q1) What is the user awareness of the privacy harms and risks that arise from them conversing to chatbots?

(Q2) What are the typical privacy-preserving behaviors practiced by chatbot users?

(Q3) How do those privacy-preserving behaviors potentially relate to the awareness of privacy threats in situations where no immediate privacy threat is perceived?

We employed mixed methods, focusing on coding answers to define broader tendencies in a quantitative way and then more closely examining answers to identify themes in a qualitative way.


By mimicking the ``what-if'' scenario where the legal framework for ``safe'' chatbot usage is in place, the aim of the study is to investigate users' awareness of privacy risks and protection measures in human-chatbot communication, as well as their privacy-preserving behaviors in relation to this awareness.


\section{Methods and Materials}

To address our research questions, we have performed a mixed-methods study with human participants with a post-study survey on digital privacy-protection awareness and actions.

In our controlled setup, each participant was asked to perform the three following tasks via a chatbot window and online forms: (1) Converse with a chatbot (at least \textit{ten} exchanges) to answer rephrased questions from a script which was not visible to the user;  (2) Answer \textit{ten} script questions (same as in 1.) in the Microsoft Forms; (3) Answer the post-study questionnaire with \textit{five} questions on privacy risk awareness.
The first and second instruments contrast the usage of chatbots and questionnaires to address Q2, the third addresses Q1, the interconnection of instruments one and three is studied to answer Q3.

Each task for instruments one and two was on one of the following three topics: (1) Your pet (with possibility to change to another topic if a person does not have a pet); (2) Your graduation ceremony; (3) Your morning routine. Topics were randomly assigned to participants under the condition the topics for instruments one and two are not the same.

\subsection{Topic Selection}

The choice of topics (Morning Routine, Graduation and Pets) represents a diverse selection of scenarios with questions (e.g., on pet's name, university a person graduated from, morning routine, etc.) and anticipated answers that are designed to elicit different types of sensitive information. In particular, we consider personal information, i.e. identifiers that can directly or indirectly, via recombination, disclose an individual, such as name or ID number, and gender and age.


Answers on the morning routine questions are anticipated to contain a lot of personal facts rather than personal identifiers. 

Answers on the graduation ceremony are expected to contain mainly indirect identifiers (graduation year and university a person graduated from). \red{While graduation information is not inherently sensitive and is often publicly available on platforms like LinkedIn, it can act as an indirect identifier. This information could be used in linkage attacks, where publicly available data is combined to identify individuals who wish to remain anonymous. For example, sharing a graduation year and university in a chatbot conversation could allow an adversary to connect this to a participant's public profile, potentially compromising their anonymity. }

Answers on the question regarding someone's pet are supposed to contain the name of the pet which is a common password (direct identifier). \red{The use of a pet's name as a password reminder question is quite common in many security settings. Many online systems offer "What is your pet's name?" as a security question for account recovery, which poses a risk since pet names are often shared on social media or in casual conversations. }

The comprehensive list of all the questions can be found in Table~\ref{scenarious}.

\begin{table}[h]
    \centering
        \caption{Table lists the ten questions designed for each of the three study topics—Pets, Graduation Ceremony, and Morning Routine. These questions were used in two formats: chatbot interactions and online questionnaires. They aimed to elicit diverse personal details, ranging from direct identifiers (e.g., pet names) to indirect identifiers (e.g., university and graduation year), providing a foundation for studying privacy awareness.}
\scalebox{.8}{
\begin{tabular}{|p{0.5cm}|l|}
\hline
\multicolumn{2}{|c|}{\textbf{\hbox{Pet}}} \\
\hline
1. & What is the name of your pet? \\
2. & What type of animal is your pet? \\
3. & What breed is your pet? \\
4. & How old is your pet? \\
5. & What is your pet's favorite food? \\
6. & Does your pet have any favorite toys or activities? \\
7. & Does your pet have any unique markings or characteristics? \\
8. & How did you first come to have your pet? \\
9. & Does your pet have any special training or tricks? \\
10. & Has your pet ever had any health issues or required medical attention? \\
\hline
\hline
\multicolumn{2}{|c|}{\textbf{\hbox{Graduation Ceremony}}} \\
\hline
1. & What university did you graduate most recently from? \\
2. & What degree did you earn at your graduation ceremony? \\
3. & Who were your guests at the graduation ceremony? \\
4. & Who was your graduation ceremony speaker? \\
5. & How long did the graduation ceremony last? \\
6. & Which year did you graduate? \\
7. & What was the dress code for your graduation ceremony? \\
8. & Did you receive any awards or honors during your graduation ceremony? \\
9. & How did you celebrate after the graduation ceremony? \\
10. & What was the most memorable moment of your graduation ceremony? \\
\hline
\hline
\multicolumn{2}{|c|}{\textbf{\hbox{Morning Routine}}} \\
\hline
1. & What time do you usually wake up in the morning on a weekday? \\
2. & What's the first thing you do when you wake up? \\
3. & Do you have a morning workout routine? \\
4. & What do you typically eat or drink for breakfast? \\
5. & Do you check your phone or social media first thing in the morning? \\
6. & Do you have a set morning routine or do you mix it up? \\
7. & How long does your morning routine usually take? \\
8. & Do you listen to music or podcasts while getting ready in the morning? \\
9. & Do you have any personal care or beauty/grooming routines in the morning? If so, please describe it. \\
10. & Do you have any morning habits that you find particularly helpful or energising? \\
\hline
\end{tabular}}
    \label{scenarious}
 \vspace{-5mm}
\end{table}

The post-study survey included five questions that asked participants about privacy risk awareness and measures they take to protect themselves in the digital world in general and while conversing with chatbots in particular (see the Post-Study Questionnaire section below).

\subsection{Participants}

We recruited 45 participants in the time period between Sep 2024 and April 2024. \red{The sample size was motivated by similar studies conducting semi-structured interviews on chatbot privacy~\citep{Gumusel24,Zhang2024}.} Participants were identified and recruited using convenience and snowballing techniques. \red{Participants were not compensated for their participation, and inclusion/exclusion criteria were based on 18+, active Internet users without any demographic restrictions. The experiment was conducted asynchronously and remotely over a six-month period.}

Four answers were discarded, which resulted in 41 participants.\footnote{Three answers were discarded from the study due to incompleteness (one of age 18-24; one of age 25-29; one unknown age group).  One participant refused to take part in the study as they felt uncomfortable with the study. } Participants were mostly university students (41\% participants of age 18-24; 15\% participants of age 25-29) and university staff (20\% participants of age 30-39, 12\% participants of age 40-49), as well as university alumni (12\% participants of age 50+). We did not collect any further demographic information but we note that the majority of participants were males. This gave a representation of an audience actively using chatbots in their everyday life. 

    \subsection{Ethics and Experimental Setup}

Study was approved by the Devolved School Research Ethics Committee at School of Electronic Engineering and Computer Science of Queen Mary University of London (QMERC20.565.DSEECS23.038). Participants were provided with the Consent Form and Information Sheet and given enough time to familiarise themselves in advance of the study. Participants could withdraw from the study at any point in time prior to first publication without needing to provide a reason. \red{Participants were provided with the Consent Form and Information Sheet and given enough time to familiarise themselves in advance of the study. Participants could withdraw from the study at any point in time prior to the first publication without needing to provide a reason. This document clearly outlined the aims of the study, the procedure of the experiment and the nature of the data being collected.}

\red{We disclosed that we aim to assess participants' awareness of privacy risks while interacting with chatbot AI. We informed our participants that they would engage in a chatbot conversation and complete an online form in the order of their choice. A post-study questionnaire will ask questions about privacy awareness and is to be completed at the end.}

\red{We specified that participants might be asked to share personal facts on topics such as their pet, graduation ceremony, or morning routine. The information sheet emphasised that providing this information was optional, with participants free to answer “I prefer not to say,'' if they felt at privacy risk or did not feel comfortable doing so.}

\red{The document assured the participants that any shared information would be fully anonymised. Additionally, participants were informed that only the act of sharing certain types of information, rather than the details themselves, was relevant to the research. The data, including chatbot conversation histories, would be stored locally, not shared with external AI providers, and further anonymised before being accessed only by the named researchers.
} 

\begin{figure}[h]
  \centering
  \includegraphics[width=0.7\hsize]{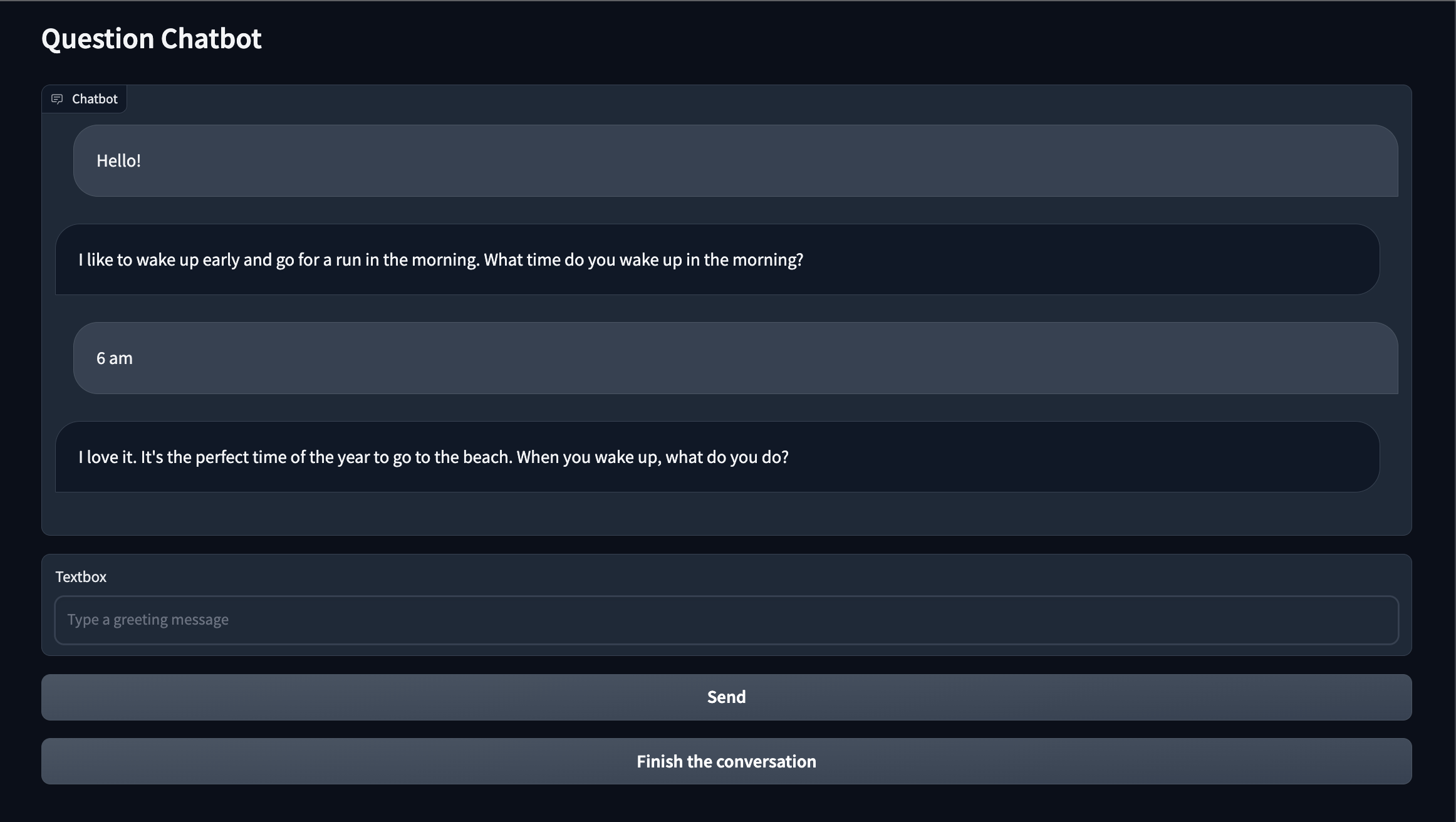}
  \caption{In-house Chatbot Interface}
  \label{fig:chatbot}
  \vspace{-5mm}
\end{figure}

Each participant was given a form and a link to the chatbot interface with questions on different topics. The chatbot used two local conversational AI models as provided by the HuggingFace library~\citep{wolf-etal-2020-transformers}: \texttt{tuner007/pegasus\_paraphrase} for creating variations of topic questions and \texttt{facebook/blenderbot-400M-distill} for the generation of chatbot comments to user questions (Figure \ref{fig:chatbot}). At no time were participant responses shared with any external AI provider. At the end, each participant was asked to complete the privacy awareness questionnaire. 

All the textual information shared via forms and the conversation with AI was pseudo-anonymised.

\subsection{Data Analysis}

\begin{figure}[h]
      \centering
      \includegraphics[width=0.8\hsize]{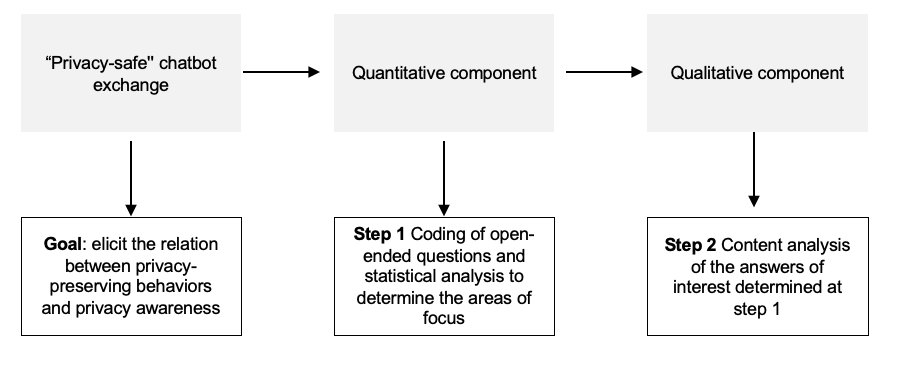}
      \caption{Study Flow Chart}
      \label{fig:methods-chart}
      \vspace{-5mm}
    \end{figure}

The flowchart of our mixed methods is in Figure~\ref{fig:methods-chart}.

At the first stage, we used quantitative methods. We started with coding of topic and post-study questions. This was an iterative
process, which involved discussions of the equally contributing authors. Discussions were repeated after coding answers of each ten participants (see Data Coding below for the final scheme). 

In addition to this informativeness score, we also measured the length of the received responses in characters and counted the number of pseudo-anonymised identifiers. \red{Together, these metrics reveal how users manage their information-sharing behaviours in chatbot interactions, shedding light on their privacy-preserving practices, and potentially reflecting their awareness of privacy risks (Q1, Q2).}

\red{We also evaluated the fluency of chatbot conversations, noting that following the script sometimes required switching topics from the previous user utterance. This transition occasionally resulted in non-fluent dialogue. Utterances were scored as 0 for non-fluent (e.g., grammatically or contextually incoherent) and 1 for fluent responses.}

For the post-study privacy awareness questionnaire, we report the coded scores per question.

This small scale quantitative analysis has allowed to define the focus of the subsequent qualitative Content Analysis. For this, the first author has manually examined the answers to identify and categorise distinct concepts. We then counted their frequencies to quantify the prevalence of certain concerns. \red{This analysis mainly examined the relationship between privacy-preserving behaviors and privacy awareness (Q3).}

We believe that this mixed methodology could be particularly useful for other researchers. The detailed quantitative analysis at the first step enhances the understanding of general trends and  allows to more efficiently focus efforts at the subsequent more costly qualitative stage.
    
\subsection{Data Coding}

We coded the textual information from the participant answers as specified below. \red{The coding system was introduced to provide more quantifiable insights into the content of responses, moving beyond the simplistic measure of response length. The length of responses alone will not be revealing enough to describe the quantity of the information shared.}

\subsubsection{Topic Questionnaire}

Final coding of responses from topic-based questionnaires involved assigning them one of the following scores:

\noindent

\textbf{Score 0}: Responses where participants refrained from disclosing any sensitive information related to the given topic (including such answers as ``I prefer not to say'' and ``I do not know'').

\textbf{Score 1}: Responses where participants provided exact answers to the questions without additional details.

\textbf{Score 1.5}: Responses with slightly more than bare minimum information. \red{This score was necessary to capture responses that provided slightly more than the bare minimum information. These responses were more detailed to be considered depersonalised short answers typical of a questionnaire but not elaborate enough to qualify as oversharing: for example, ``no, I do not think so'', ``no, not at all''.} 

\textbf{Score 2}: Responses where participants shared more information than necessary, revealing additional details beyond what was explicitly asked.

\subsubsection{Post-Study Survey}

Coding responses from post-study questionnaires involved the following:

\textit{Q1 (categorical): Which of the following privacy risks are you aware of (a) disclosing PII (Personally
Identifiable Information), (b) risk of being identified from the personal writing style and
(c) risk of being profiled for personality type?} Answers were assigned a numerical score based on the number of options selected.

\textit{Q2 (open-ended): What steps do you normally take to protect your privacy online?} As there is no comprehensive list of standard procedures for online privacy protection, the responses were examined for practices grouped into technology-based measures (e.g., using VPN, managing cookies) and self-organisation measures (e.g., changing personal information, setting complex passwords). The score was hence 0, 1, or 2 depending on how many types of these measures were mentioned.

\textit{Q3 (open-ended): Why sharing the information with a chatbot can pose a privacy risk?} Possible reasons mentioned by participants in the answers involved ambiguity in handling data and regulation issues. We assigned a score from 0 to 2 based on the presence of these points.

\textit{Q4 (open-ended): How often do you converse with chatbots?} A binary scoring system was employed based on participants' reported frequency of chatbot usage. Those who reported using the chatbot at least once a week received a score of 1, while others received a score of 0.

\textit{Q5 (open-ended): If you do, do you take any steps to protect your privacy?} We employed scoring based on the reported actions or lack thereof: 0 score for no actions, 1 for a limited number of actions, 2 for two and more actions.

\section{Results}

Tables below presents the results of our analysis.

\begin{table}[h]
    \centering
        \caption{Quantitative comparison of participant responses between the questionnaire and chatbot across three scenarios (Morning, Graduation, and Pet). Columns include Informativeness, representing the informativeness score (mean ± std dev), Length in Chars, the average response length (in characters), Fluency, the fluency score for chatbot responses (mean ± std dev), and PIs, the count of pseudo-anonymised identifiers (mean ± std dev). The chatbot conversations show statistically significant increase in informativeness scores (denoted by *, measured by pairwise t-test). Scores are representative and vary across topics.}
\scalebox{.8}{
\begin{tabular}{cccc|@{\hspace{1cm}}cccc}
 & \multicolumn{3}{c}{Questionnaire} 
& \multicolumn{4}{c}{Chatbot}  \\ 
\cline{2-8}
 & Informativeness & Length in Chars & PIs & Informativeness & Length in Chars & Fluency & PIs \\
\midrule
Morning & 13.0 $\pm$ 2.7 & 363$\pm$326 & 0.0 $\pm$ 0.0 & 14.0 $\pm$ 3.0 & 457$\pm$222 & 7.6 $\pm$ 1.3 & 1.9 $\pm$ 1.7 \\ 
Graduation & 11.0 $\pm$ 2.7 & 219 $\pm$112 & 2.0 $\pm$ 0.4 & 15.0 $\pm$ 4.3 & 396$\pm$225 & 7.4 $\pm$ 2.4 & 0.6 $\pm$ 1.5 \\
Pet & 12.0 $\pm$ 1.3 & 178$\pm$70 & 0.9 $\pm$ 0.4 & 15.0 $\pm$ 3.6 & 426$\pm$253 & 7.3 $\pm$ 2.3 & 1.4 $\pm$ 2.0 \\
\midrule
Overall & 12.0 $\pm$ 2.4 & 254$\pm$216 & 0.9 $\pm$ 0.9 & 15.0* $\pm$ 3.6 & 396$\pm$225 & 7.4 $\pm$ 2.0 & 1.3 $\pm$ 1.8\\
  \bottomrule
\end{tabular}}
    \label{general-stat}
 \vspace{-5mm}
\end{table}

\begin{table}[h]
    \centering
        \caption{Quantitative comparison of participant responses for the post-study privacy awareness survey. The table presents average scores (Av) with standard deviations (± std dev) for each question (Q1 to Q5) and the distribution of scores across response categories (0, 1, 2, 3). Q1 measures risk awareness and Q2 evaluates privacy-preserving measures. For Chatbot Usage, Q3 assesses awareness of risks, Q4 captures regularity of chatbot interactions and Q5 explores measures taken to ensure privacy during chatbot use. Participants were aware of an average of two privacy risks and typically employed only one privacy-preserving measure, with 63\% conversing with chatbots at least weekly and only 24\% relying on some strategies to protect their privacy during those conversations.}
\scalebox{.8}{        
\begin{tabular}{ccc|ccc|c}
& \multicolumn{2}{c}{Overall}  & \multicolumn{3}{c}{Chatbot usage} & \\ \toprule
& Q1: risk awareness & Q2: measures & Q3: risks & Q4: regularity & Q5: measures & Av \\
\midrule
Av & 2.0 $\pm$ 0.9 & 1.3 $\pm$ 0.9 & 1.4 $\pm$ 0.6 & 0.7 $\pm$ 0.5  & 0.9 $\pm$ 0.8 & 1.3 $\pm$ 0.4 \\
\midrule
0, \% & 5 &  10 & 7 & 37 &  32 & - \\
1, \% & 24 & 51 & 41 & 63 & 44 & - \\
2, \% & 37 & 39 & 52 & - & 24  & - \\
3, \% & 34 & - & - & - &  - & - \\
  \bottomrule
\end{tabular}}
    \label{post-stat}
 \vspace{-5mm}
\end{table}

\begin{table}[h]
    \centering
        \caption{Size in participants (\#) and statistics of answers per age group. We report the average score of the quantity of shared information (Informativeness), Response length in characters (Length in Chars), count of personal identifiers (PIs, \#), as well as utterance fluency (Fluency) .Comparison of participant responses across age groups in the Questionnaire and Chatbot responses. For each age group, the table presents average scores (mean ± std dev) for informativeness, response length (in characters), fluency (for chatbot responses), and the number of pseudo-anonymised identifiers (PIs). The chatbot conversations are more informative than the questionnaire responses within both 18-24 and 25+ groups (statistically significant denoted by *, measured by pairwise t-test).}
\scalebox{.8}{
\begin{tabular}{ccccc|@{\hspace{1cm}}cccc}
&  & \multicolumn{3}{c}{Questionnaire} 
& \multicolumn{4}{c}{Chatbot} \\ \midrule
& \# & Informativeness & Length in Chars & PIs & Informativeness & Length in Chars & Fluency & PIs\\
\midrule
18-24 & 17 & 12.2$\pm$2.3 & 292$\pm$246 & 0.8$\pm$1.0 & 14.8$\pm$3.2* & 427$\pm$257 & 6.9$\pm$2.2 & 1.2$\pm$1.1 \\ 
\midrule
25+ & 24 & 11.6$\pm$2.6 & 227$\pm$193 & 1.0$\pm$0.8 & 15.1$\pm$3.9* & 425$\pm$214 & 7.8$\pm$1.8 & 1.4$\pm$2.1 \\
\midrule
\midrule
25-29 & 6 & 12.3$\pm$2.5 & 304$\pm$287 & 1.0$\pm$0.9 & 15.7$\pm$3.9 & 416$\pm$223 & 8.3$\pm$2.1 & 0.7$\pm$1.0 \\
30-39 & 8 & 13.0$\pm$1.9 & 288$\pm$196 & 1.1$\pm$0.8 & 16.0$\pm$2.3 & 418$\pm$165  & 7.8$\pm$2.1 & 1.0$\pm$1.1 \\
40-49 & 5 & 10.1$\pm$0.2 & 121$\pm$35 & 0.6$\pm$0.9 & 11.9$\pm$1.6 & 221$\pm$54 & 7.6$\pm$1.7 & 0.4$\pm$0.9 \\
50+ & 5 & 10.2$\pm$3.6 & 142$\pm$60 & 1.2$\pm$0.8 & 16.2$\pm$6.3 & 649$\pm$191 & 7.2$\pm$1.3 & 3.8$\pm$3.6  \\
  \bottomrule
\end{tabular}}
    \label{per-age-stat}
 \vspace{-5mm}
\end{table}

\begin{table}[h]
    \centering
        \caption{Size in participants (\# ) and statistics of answers per age group. For the post-study privacy awareness survey, we report the scores per question and the average score by age group. The table presents mean scores (± standard deviation) for each question. Participants from the 18-24 cohort scored lower overall (Av: 1.2 ± 0.4) compared to those 25+ (Av: 1.3 ± 0.4), with notable differences in Q4 (regularity of chatbot interactions, statistically significant denoted by *, measured by pairwise t-test). Among subgroups, the privacy-awareness increased with age (see the growth of the average score from group 25-29 to group 50+).}
\scalebox{.8}{
\begin{tabular}{cccc|@{\hspace{1cm}}ccc|c}
&  & \multicolumn{2}{c}{Overall} & \multicolumn{3}{c}{Chatbot usage} & \\ \toprule
& \# & Q1: risk awareness & Q2: measures & Q3: risks & Q4: regularity & Q5: measures & Av \\
\midrule
18-24 & 17 & 1.8$\pm$0.8 & 1.3$\pm$0.8 & 1.5$\pm$0.7 & 0.5$\pm$0.5 & 0.8$\pm$0.8 & 1.2$\pm$0.4 \\ 
\midrule
25+ & 24 & 2.1$\pm$0.9 & 1.3$\pm$0.6 & 1.4$\pm$0.6 & 0.8$\pm$0.5* & 1.0$\pm$0.7 & 1.3$\pm$0.4 \\
\midrule
\midrule
25-29 & 6 & 2.3$\pm$1.0 & 1.0$\pm$0.6 & 1.2$\pm$0.8 & 0.8$\pm$0.4 & 0.7$\pm$0.8 & 1.2$\pm$0.5 \\
30-39 & 8 & 2.1$\pm$0.6 & 1.6$\pm$0.7 & 1.5$\pm$0.5 & 0.5$\pm$0.5 & 0.9$\pm$0.6 & 1.3$\pm$0.3 \\
40-49 & 5 & 2.0$\pm$1.4 & 1.2$\pm$0.4 & 1.2$\pm$0.4 & 1.2$\pm$0.4 & 1.2$\pm$0.8 & 1.4$\pm$0.5 \\
50+ & 5 & 2.0$\pm$1.0 & 1.4$\pm$0.5 & 1.8$\pm$0.4 & 0.8$\pm$0.4 & 1.4$\pm$0.5 & 1.5$\pm$0.2 \\
  \bottomrule
\end{tabular}}
    \label{per-age-post-stat}
 \vspace{-5mm}
\end{table}

\subsection{Quantitative Analysis}

The general tendency in our results supports the trend from the recent literature about the oversharing of information in chatbot conversations~\cite{Pizzi2023}. Table~\ref{general-stat} shows that the answers to the bot question receive slightly higher informativeness scores than those of the questionnaires (\red{cf. the statistically significant difference in average Informativeness in chatbot answers as compared to questionnaires}). \red{The standard deviation for chatbots is also higher}. User utterances in chatbot conversations are in general twice as long as questionnaire answers (\red{cf. the average Length in Chars for Questionnaire and Chatbot}). Also, the average count of direct or indirect identifiers found in the chatbot conversations is slightly higher than this count for the questionnaire (\red{cf. average PI counts for Questionnaire and Chatbot}). 

The selection of topics in the study enabled a representative range of characteristics to be covered. For instance, average informativeness scores remained stable across topics for both questionnaires and chatbot conversations. Characteristics such as response length and the count of identifiers showed a range of representative values across topics, e.g., the average count of identifiers in questionnaire answers was 0, 2 and 0.9, for morning, graduation and pet respectively (\red{see the PI column for Questionnaire in~Table~\ref{general-stat}}). The tendency for higher information disclosure in chatbot conversations was consistent across all topics.

The participants were aware of two privacy risks on average (\red{see average score for Q1: risk awareness, Table~\ref{post-stat}}). Usually, only one type of measures were taken to protect privacy online (\red{see average score for Q2: measures)}), it was predominantly a technology-based measure (such as private browsing, being selective to cookies). 

The major concern of our participants was that they were unaware of who was accessing their data. 63\%  of respondents conversed with chatbots at least once a week (\red{see distribution for Q3: risks and Q4: regularity for Chatbot usage}). Only 24\% of our respondents took more than one measure to protect their privacy while conversing (\red{see distribution for Q5: measures}). This was mostly not revealing PII, avoiding talking on personal topics or rephrasing their answers. 

Table~\ref{per-age-stat} shows that our observations are also consistent across age groups: the averages for our well-represented cohort of 18-24 overall matches the averages of the cohort 25+, both in terms of response informativeness, as well as the privacy awareness scores (\red{cf. rows 18-24 and 25+}). Inside the cohort 25+ we observe some fluctuations in sub-cohorts 40-49 and 50+. We believe that those are due to non-representative samples in those smallest age groups (\red{this is due to the limited number of participants with predominantly with STEM backgrounds}). 

\red{Results in Table~\ref{per-age-post-stat} revealed a statistically significant difference between the 25+ age group and the 18–24 group in terms of chatbot interactions, with the 25+ group scoring higher (cf. Q4: regularity of chatbot interactions) for rows 18-24 and 25+). It is possible that younger participants tend to conceal their chatbot usage, as many are still studying and avoid disclosing this usage as negatively perceived in the education. Overall, participant awareness of privacy risks grew with age (see steady increase in the average score from age 25-29 to age 50+)}

Overall, our conclusions indicate a partial awareness of online privacy risks, while awareness regarding communication with chatbots remains limited. This is concerning given the frequency of daily interactions and the tendency to overshare personal information evidenced above. This situation calls for more proactive measures to enhance privacy awareness among chatbot users, such as mandatory training in workplaces or specialised tutorials for personal use. 

In the absence of dedicated tools for privacy protection in chatbot communication, common privacy-preserving behaviors typically involve cognitively demanding activities like avoiding certain topics or rephrasing personal information~\citep{Zhang2024}. Automating these activities could help reduce cognitive load but might also increase the time and frustration associated with communication. Therefore, we conducted a more in-depth study of the chatbot conversations in our qualitative analysis to understand the potential trade-offs of implementing such tools. 

\subsection{Qualitative Analysis}

Our initial analysis has highlighted post-study questions 3 and 5 as the most relevant for further examination. Consequently, we conducted content analysis of the chatbot conversations, as well as these questions to address our research inquiries as follows:

(Q1) What is the user awareness of the privacy harms and risks that arise from them conversing to chatbots?

\textbf{Idea ``It will store and use my data but we do not know how''}

The first remark is that 27\% (11 participants) mention that chatbots store and use data (see Table S1 in Appendix B). However, the exact procedures or consequences are mostly unclear to the participants: only six participants mention identity disclosure; another six participants name concrete scenarios that might follow a privacy breach, e.g., blackmailing, unwanted emails or ads, tracking, etc.; finally four participants mention data leaks. The fact that users rarely mention any consequences of privacy harms indicate low awareness of those consequences. Our conclusion is supported by the comment of one participant who mentioned that they do not understand how their data could be used for malicious purposes. 

``If someone gained access to the messages I sent to the chatbot, they could uncover my identity. But I am not sure if this information could be of any use to harm me.''\footnote{Hereinafter, to preserve privacy, all the examples are created by ChatGPT (\url{https://chatgpt.com/}) and motivated by the real user answers in the study.}

The ambiguity surrounding the chatbot's data handling procedures is further highlighted by participants' experiences: six individuals felt the chatbot was personified, while four perceived the company as a malicious entity manipulating the data.

``It (chatbot) stores the information about me without my consent. It makes it easier for others to exploit me through targeted ads or even blackmail.''

Overall, only two answers mentioned paying attention to the reliability of the chatbot provider (fact that there could be trusted and not trusted chatbots) indicating the lack of knowledge on relevant legal regulations. And only one participant has mentioned the lack of regulations regarding the usage of the chatbot-collected data. 

\textbf{Idea ``They will clone me''}

Other slightly less popular tendencies include chatbot ``learning'' user profiles to further ``exploit'' them (three participants):

``The chatbot can delve into your messages and uncover the essence of who you are.''

\textbf{Idea ``Our data will end on the black market''}

Another interesting fear is the fear of participants that their data will be sold (three participants):

``We have no idea what happens to our information after we talk to a chatbot. They could be selling our data, and we are left in the dark about how it's being used.''

Comparing the content of the answers across age groups, we have noticed that all the three participants are from the 25+ cohort.

(Q2) What are the typical privacy-preserving behaviors practiced by chatbot users?

We need to emphasise that generic measures to protect privacy online were better known and more widely used by our participants: 90\% of participants (37 participants) mentioned some measures as private browsing, VPNs, complex passwords, temporary email accounts, avoiding PII, vigilance while accepting cookies, etc. (see Table~\ref{post-stat}) While regarding the privacy protection measures while conversing with chatbots 29\% of our participants (12 participants) took no measures. Nine out of twelve are under thirty. 

\textbf{Idea ``Someone I know was wondering...''}

Beyond inactivity, one of the most prominent strategies was the strategy not to mention Personally Identifiable Information (PII). It was mentioned in~27\% of cases (11 participants, see Table S2 in Appendix B). 

Other strategies included trying to minimise the information while conversing with chatbots (mentioned by nine participants), as well as the tendency to conceal personality (three participants). This was done by intentionally altering personal data or generalising, or just avoiding conversations on personal topics (beliefs or ideology).  

Privacy-preserving behaviours in chatbot conversations were more typical for the 25+ demographic group, e.g., from all the individuals seeking to protect PII only two out of eleven were from the 18-24 group.



(Q3) How do those privacy-preserving behaviors potentially relate to the awareness of privacy threats in situations where no immediate privacy threat is perceived?

To answer this question we have mainly been focusing on the two most \red{contrasting} privacy-preserving behaviours in chatbot conversations: inactivity and attempts to minimise the shared information or conceal personality.

\textbf{Idea ``Not interested''}

For the ``do nothing'' responses we have mostly noticed the lack of engagement and/or conveying any particular importance to information disclose. Hence, the answers were partially very short, often containing identifiers or the phrase ``I prefer not to say''. See Appendix A in Supplementary material for examples of such answers contrasted to more detailed examples. 

Regarding the over-protective behaviors (``minimising information'' and ``concealing personality''), 30\% of the answers (4 answers) were rather open and natural. In three cases dialogues mixed emotional open answers with deflective behavior (e.g., answering the question with an over protective question, see Appendix B in the Supplementary material). We have noticed that such answers were often associated with pseudo-professional technical or legal knowledge of privacy which hints towards the ``privacy paradox''. Only 42\% of the answers (5 answers) contained short dry utterances, actively concealing personality.

\red{Overall, our qualitative analysis highlights a general lack of trust in chatbot interactions. However, in trusted environments, participants, particularly younger individuals and domain experts, often forgot privacy-protection measures, suggesting that such behaviors do not come naturally. This aligns with previous findings on disclosure tendencies influenced by trust in the recipient of shared information~\citep{Gerber2018}.} 

This  further indicates that concealing private information may not come naturally, especially for individuals with privacy expertise. This opens pathways of discussion around the privacy paradox. 

\red{People with extensive expertise may overshare for several reasons: experts often prioritise functionality over privacy concerns. Their confidence in managing technology can lead to overestimating their ability to control risks (e.g., knowledge of the time needed to find the right combination of identifiers, having an idea about the amount of noise in the information flow and the difficulty to capture  relevant information in it), coupled with a sense of trust in the system or environment~\citep{Barth2019}. Finally, the reason may be ``privacy fatigue'' due to comprehensive knowledge of potential privacy issues.} 

Therefore, implementing a mechanism for reminders could be beneficial to promote more consistent privacy-preserving behavior in certain scenarios with individuals with in-depth expertise in AI and its Ethics.
 

\section{Discussion}


Our study investigated the relation between privacy awareness and privacy-preserving behaviors in a controlled ``privacy-safe" environment, mimicking the ``what-if'' scenario of legally protected chatbot communication. We found a significant lack of understanding among participants about data handling by chatbots, leading to mistrust. We have also found inconsistencies in user behaviours regarding privacy: they did not consistently exhibit privacy-preserving behaviors, possibly due to over-reliance on technical protections.

Our study confirms findings in the existing literature that users have limited understanding of how chatbots process and store data, contributing to mistrust. It also enhances the notion of privacy paradox, where users express concerns but do not consistently protect their privacy, which we investigated in the unique ``privacy-safe settings''.

\subsection{Study Strengths and Limitations}

One of the main strengths and novelties of our study is the controlled environment that ensured participant privacy, allowing us to focus on behaviors without the confounding influence of privacy breaches. This methodology can be adapted to different scenarios to investigate privacy-preserving behaviours. The use of mixed methods (both quantitative and qualitative) enables more granular understanding of the awareness of the related privacy issues.

Our study's vast scope makes it inherently limited due to the sample size and the fact that we were limited to academic circles in Computer Science. Additionally, the limited representation of women in Computer Science further constrain the applicability of our results. Nevertheless, we believe that our participants, who work closely with AI, mostly represent an upper bound of chatbot privacy awareness.

\subsection{Implications for Research, Practice \& Policy}

For research, our findings highlight the paths to further explore the privacy paradox and develop better educational tools on data privacy risks such as identity theft and manipulation. In practice, comprehensive data privacy courses should be made widely available. Additionally, a strategic approach might be needed to adjust the over- or under-protective user behaviors, such as developing tools that help users monitor and control the information they share with chatbots. In our case, under-protective behaviors were observed among more expert users.

Our findings also offer valuable contributions for policymakers, aiding in the development of policy objectives related to privacy risk awareness across demographic groups and data sharing patterns. They also contribute to the design of tools that inform privacy policies on action mechanisms and success measures.

\red{This article’s relevance to public health lies in its focus on privacy concerns associated with chatbot interactions, which are increasingly used in areas such as mental health support, patient education, and symptom tracking~\citep{wilson}. While GOV.UK Digital Services \footnote{\url{https://www.gov.uk/guidance/using-chatbots-and-webchat-tools}} highlights the importance of transparency and data protection compliance, it does not provide practical methods on how to increase user trust.  Understanding user awareness and behaviors regarding privacy risks is critical for designing chatbots that protect sensitive health information while maintaining user trust. }

\section{Conclusion}

Our study investigates human privacy-preserving behaviors in a controlled ``privacy-safe'' environment using an ad-hoc chatbot. Our goal was to better understand user awareness of privacy risks in chatbot conversations, their typical privacy-preserving behaviors, and how these behaviors relate to privacy awareness in trusted conditions. Confirming previous studies, we identified a lack of understanding regarding how chatbots handle data and the implications of data breaches. Users, particularly those with privacy expertise, do not consistently exhibit privacy-preserving behaviors when assured of the chatbot's transparent data processing procedures. This inconsistency may be due to an over-reliance on technical protection measures. One dialogue can display both very protective and overly open tendencies, opening pathways for further research within the privacy paradox sphere where users express concerns about privacy but do not consistently act to protect it.

\section*{Conflict of Interest Statement}

The authors declare that the research was conducted in the absence of any commercial or financial relationships that could be construed as a potential conflict of interest.

\section*{Author Contributions}

JI conceived the approach, conducted the experiments and wrote the paper. JI and VY coded the transcripts and analysed the data. MI, MR and PB assisted in writing and review. All reviewed the research and the manuscript. All authors approved the manuscript.

\section*{Funding}
No funding was received for this work.

\section*{Acknowledgments}

\red{We acknowledge the help of ChatGPT 4o to edit the manuscript, the content edited using the Generative AI has been checked for factual accuracy and plagiarism.}
 
\section*{Data Availability Statement}
\red{Data cannot be publicly shared due to confidentiality requirements; however, access can be granted upon obtaining relevant ethics clearance.}  

\bibliographystyle{Frontiers-Harvard} 
\bibliography{test}

\begin{thebibliography}{28}
\providecommand{\natexlab}[1]{#1}
\expandafter\ifx\csname urlstyle\endcsname\relax
  \providecommand{\doi}[1]{doi:\discretionary{}{}{}#1}\else
  \providecommand{\doi}{doi:\discretionary{}{}{}\begingroup \urlstyle{rm}\Url}\fi
\providecommand{\selectlanguage}[1]{\relax}
\providecommand{\bibAnnoteFile}[1]{%
  \IfFileExists{#1}{\begin{quotation}\noindent\textsc{Key:} #1\\
  \textsc{Annotation:}\ \input{#1}\end{quotation}}{}}
\providecommand{\bibAnnote}[2]{%
  \begin{quotation}\noindent\textsc{Key:} #1\\
  \textsc{Annotation:}\ #2\end{quotation}}

\bibitem[{Araujo and Bol(2024)}]{Araujo2024}
Araujo, T. and Bol, N. (2024).
\newblock From speaking like a person to being personal: The effects of personalized, regular interactions with conversational agents.
\newblock \emph{Computers in Human Behavior: Artificial Humans} 2, 100030.
\newblock \doi{10.1016/J.CHBAH.2023.100030}
\bibAnnoteFile{Araujo2024}

\bibitem[{Barth et~al.(2019)Barth, de~Jong, Junger, Hartel, and Roppelt}]{Barth2019}
Barth, S., de~Jong, M.~D., Junger, M., Hartel, P.~H., and Roppelt, J.~C. (2019).
\newblock Putting the privacy paradox to the test: Online privacy and security behaviors among users with technical knowledge, privacy awareness, and financial resources.
\newblock \emph{Telematics and Informatics} 41, 55--69.
\newblock \doi{10.1016/j.tele.2019.03.003}
\bibAnnoteFile{Barth2019}

\bibitem[{Colnago et~al.(2023)Colnago, Cranor, and Acquisti}]{Colnago2023}
Colnago, J., Cranor, L.~F., and Acquisti, A. (2023).
\newblock Is there a reverse privacy paradox? an exploratory analysis of gaps between privacy perspectives and privacy-seeking behaviors; is there a reverse privacy paradox? an exploratory analysis of gaps between privacy perspectives and privacy-seeking behaviors \doi{10.56553/popets-2023-0027}
\bibAnnoteFile{Colnago2023}

\bibitem[{Dev and Dev(2023)}]{Dev2023}
Dev, J. and Dev, S. (2023).
\newblock "how can i help you?": User perceptions of privacy in retail chat agents.
\newblock \emph{Conference on Human Factors in Computing Systems - Proceedings} \doi{10.1145/3544549.3585796}
\bibAnnoteFile{Dev2023}

\bibitem[{Gerber et~al.(2018)Gerber, Gerber, and Volkamer}]{Gerber2018}
[Dataset] Gerber, N., Gerber, P., and Volkamer, M. (2018).
\newblock Explaining the privacy paradox: A systematic review of literature investigating privacy attitude and behavior.
\newblock \doi{10.1016/j.cose.2018.04.002}
\bibAnnoteFile{Gerber2018}

\bibitem[{Gumusel et~al.(2024)Gumusel, Zhou, and Sanfilippo}]{Gumusel24}
Gumusel, E., Zhou, K.~Z., and Sanfilippo, M.~R. (2024).
\newblock User privacy harms and risks in conversational ai: A proposed framework
\bibAnnoteFile{Gumusel24}

\bibitem[{Ischen et~al.(2020)Ischen, Araujo, Voorveld, van Noort, and Smit}]{Ischen2020}
Ischen, C., Araujo, T., Voorveld, H., van Noort, G., and Smit, E. (2020).
\newblock Privacy concerns in chatbot interactions.
\newblock \emph{Lecture Notes in Computer Science (including subseries Lecture Notes in Artificial Intelligence and Lecture Notes in Bioinformatics)} 11970 LNCS, 34--48.
\newblock \doi{10.1007/978-3-030-39540-7_3}
\bibAnnoteFile{Ischen2020}

\bibitem[{Kokolakis(2017)}]{Kokolakis2017}
Kokolakis, S. (2017).
\newblock Privacy attitudes and privacy behaviour: A review of current research on the privacy paradox phenomenon.
\newblock \emph{Computers \& Security} 64, 122--134.
\newblock \doi{10.1016/J.COSE.2015.07.002}
\bibAnnoteFile{Kokolakis2017}

\bibitem[{Lappeman et~al.(2023)Lappeman, Marlie, Johnson, and Poggenpoel}]{Lappeman2023}
Lappeman, J., Marlie, S., Johnson, T., and Poggenpoel, S. (2023).
\newblock Trust and digital privacy: willingness to disclose personal information to banking chatbot services.
\newblock \emph{Journal of Financial Services Marketing} 28, 337--357.
\newblock \doi{10.1057/S41264-022-00154-Z/TABLES/15}
\bibAnnoteFile{Lappeman2023}

\bibitem[{Larsen and Følstad(2024)}]{Larsen2024}
Larsen, A.~G. and Følstad, A. (2024).
\newblock The impact of chatbots on public service provision: A qualitative interview study with citizens and public service providers.
\newblock \emph{Government Information Quarterly} 41, 101927.
\newblock \doi{10.1016/J.GIQ.2024.101927}
\bibAnnoteFile{Larsen2024}

\bibitem[{Li et~al.(2023)Li, Zhang, Lee, Kraut, and Mohr}]{Li2023}
Li, H., Zhang, R., Lee, Y.~C., Kraut, R.~E., and Mohr, D.~C. (2023).
\newblock Systematic review and meta-analysis of ai-based conversational agents for promoting mental health and well-being.
\newblock \emph{npj Digital Medicine} 6.
\newblock \doi{10.1038/S41746-023-00979-5}
\bibAnnoteFile{Li2023}

\bibitem[{Liu et~al.(2024)Liu, Yao, Jia, Casper, Baracaldo, Hase et~al.}]{unlearning}
Liu, S., Yao, Y., Jia, J., Casper, S., Baracaldo, N., Hase, P., et~al. (2024).
\newblock Rethinking machine unlearning for large language models
\bibAnnoteFile{unlearning}

\bibitem[{Ma et~al.(2024)Ma, Mei, Long, Su, Bren, and Gajos}]{lgbt24}
Ma, Z., Mei, Y., Long, Y., Su, Z., Bren, D., and Gajos, K.~Z. (2024).
\newblock Evaluating the experience of lgbtq+ people using large language model based chatbots for mental health support evaluating the experience of lgbtq+ people using large language * equal contributions model based chatbots for \doi{10.1145/3613904.3642482}
\bibAnnoteFile{lgbt24}

\bibitem[{Pizzi et~al.(2023)Pizzi, Vannucci, Mazzoli, and Donvito}]{Pizzi2023}
Pizzi, G., Vannucci, V., Mazzoli, V., and Donvito, R. (2023).
\newblock I, chatbot! the impact of anthropomorphism and gaze direction on willingness to disclose personal information and behavioral intentions.
\newblock \emph{Psychology \& Marketing} 40, 1372--1387.
\newblock \doi{10.1002/MAR.21813}
\bibAnnoteFile{Pizzi2023}

\bibitem[{Rukadikar and Khandelwal(2024)}]{Rukadikar2024}
Rukadikar, A. and Khandelwal, K. (2024).
\newblock Navigating change: a qualitative exploration of chatbot adoption in recruitment.
\newblock \emph{Cogent Business \& Management} 11, 2345759.
\newblock \doi{10.1080/23311975.2024.2345759}
\bibAnnoteFile{Rukadikar2024}

\bibitem[{Shah and Panchal(2022)}]{Shah2022}
Shah, M.~H. and Panchal, M. (2022).
\newblock Theoretical evaluation of securing modules for educational chatbot.
\newblock \emph{Proceedings - 2022 6th International Conference on Intelligent Computing and Control Systems, ICICCS 2022} , 818--824\doi{10.1109/ICICCS53718.2022.9788120}
\bibAnnoteFile{Shah2022}

\bibitem[{Solove(2020)}]{Solove2020}
Solove, D. (2020).
\newblock The myth of the privacy paradox.
\newblock \emph{GW Law Faculty Publications \& Other Works}
\bibAnnoteFile{Solove2020}

\bibitem[{Veale(2024)}]{veale}
Veale, M. (2024).
\newblock Rights for those who unwillingly, unknowingly and unidentifiably compute! \doi{10.31235/OSF.IO/4UGXD}
\bibAnnoteFile{veale}

\bibitem[{Völkel et~al.(2020)Völkel, Haeuslschmid, Werner, Hussmann, and Butz}]{volker20}
Völkel, S.~T., Haeuslschmid, R., Werner, A., Hussmann, H., and Butz, A. (2020).
\newblock How to trick ai: Users' strategies for protecting themselves from automatic personality assessment.
\newblock \emph{Conference on Human Factors in Computing Systems - Proceedings} \doi{10.1145/3313831.3376877}
\bibAnnoteFile{volker20}

\bibitem[{Wang et~al.(2023)Wang, Chen, Pei, Xie, Kang, Zhang et~al.}]{wang2023decodingtrust}
Wang, B., Chen, W., Pei, H., Xie, C., Kang, M., Zhang, C., et~al. (2023).
\newblock Decodingtrust: A comprehensive assessment of trustworthiness in {GPT} models.
\newblock In \emph{Thirty-seventh Conference on Neural Information Processing Systems Datasets and Benchmarks Track}
\bibAnnoteFile{wang2023decodingtrust}

\bibitem[{Wei et~al.(2023)Wei, Haghtalab, and Steinhardt}]{Wei2023JailbrokenHD}
Wei, A., Haghtalab, N., and Steinhardt, J. (2023).
\newblock Jailbroken: How does llm safety training fail?
\newblock \emph{ArXiv} abs/2307.02483
\bibAnnoteFile{Wei2023JailbrokenHD}

\bibitem[{Weizenbaum(1966)}]{Weizenbaum1966}
Weizenbaum, J. (1966).
\newblock Eliza—a computer program for the study of natural language communication between man and machine.
\newblock \emph{Communications of the ACM} 9, 36--45.
\newblock \doi{10.1145/365153.365168}
\bibAnnoteFile{Weizenbaum1966}

\bibitem[{Willems et~al.(2023)Willems, Schmid, Vanderelst, Vogel, and Ebinger}]{Willems2023}
Willems, J., Schmid, M.~J., Vanderelst, D., Vogel, D., and Ebinger, F. (2023).
\newblock Ai-driven public services and the privacy paradox: do citizens really care about their privacy?
\newblock \emph{Public Management Review} 25, 2116--2134.
\newblock \doi{10.1080/14719037.2022.2063934}
\bibAnnoteFile{Willems2023}

\bibitem[{Wilson and Marasoiu(2022)}]{wilson}
Wilson, L. and Marasoiu, M. (2022).
\newblock The development and use of chatbots in public health: Scoping review.
\newblock \emph{JMIR Hum Factors} 9, e35882.
\newblock \doi{10.2196/35882}
\bibAnnoteFile{wilson}

\bibitem[{Wolf et~al.(2020)Wolf, Debut, Sanh, Chaumond, Delangue, Moi et~al.}]{wolf-etal-2020-transformers}
Wolf, T., Debut, L., Sanh, V., Chaumond, J., Delangue, C., Moi, A., et~al. (2020).
\newblock Transformers: State-of-the-art natural language processing.
\newblock In \emph{Proceedings of the 2020 Conference on Empirical Methods in Natural Language Processing: System Demonstrations} (Online: Association for Computational Linguistics), 38--45.
\newblock \doi{10.18653/v1/2020.emnlp-demos.6}
\bibAnnoteFile{wolf-etal-2020-transformers}

\bibitem[{Wörsdörfer(2023)}]{worsdorfer24}
Wörsdörfer, M. (2023).
\newblock The e.u.’s artificial intelligence act: An ordoliberal assessment.
\newblock \emph{SSRN Electronic Journal} \doi{10.2139/SSRN.4544276}
\bibAnnoteFile{worsdorfer24}

\bibitem[{Yu et~al.(2024)Yu, Kairouz, Oh, and Xu}]{Yu2024}
Yu, D., Kairouz, P., Oh, S., and Xu, Z. (2024).
\newblock Privacy-preserving instructions for aligning large language models
\bibAnnoteFile{Yu2024}

\bibitem[{Zhang et~al.(2024)Zhang, Jia, Hao-Ping, Lee, Yao, Das et~al.}]{Zhang2024}
Zhang, Z., Jia, M., Hao-Ping, Lee, Yao, B., Das, S., et~al. (2024).
\newblock "it's a fair game'', or is it? examining how users navigate disclosure risks and benefits when using llm-based conversational agents 1
\bibAnnoteFile{Zhang2024}

\end{thebibliography}


\end{document}